\documentstyle[prb,aps,epsf,twocolumn]{revtex}
\begin{document}
\draft

\title{Density functional study of Au$_n$ ($n=2-20$) clusters: lowest-energy structures and electronic properties}

\author{Jinlan Wang$^{1^{*}}$, Guanghou Wang$^{1^{*}}$, Jijun Zhao$^2$$^{\dagger }$}
\address{$^1$National Laboratory of Solid State Microstructures and Department of physics, Nanjing University, Nanjing 210093, P.R. China \\
$^2$Department of Physics and Astronomy, University of North Carolina at Chapel Hill, Chapel Hill, North Carolina 27599-3255}
\maketitle

\begin{abstract}
We have investigated the lowest-energy structures and electronic properties of the Au$_n$($n=2-20$) clusters based on density functional theory (DFT) with local density approximation. The small Au$_n$ clusters adopt planar structures up to $n=6$. Tabular cage structures are preferred in the range of $n=10-14$ and a structural transition from tabular cage-like structure to compact near-spherical structure is found around $n=15$. The most stable configurations obtained for Au$_{13}$ and Au$_{19}$ clusters are amorphous instead of icosahedral or fcc-like, while the electronic density of states sensitively depend on the cluster geometry. Dramatic odd-even alternative behaviors are obtained in the relative stability, HOMO-LUMO gaps and ionization potentials of gold clusters. The size evolution of electronic properties is discussed and the theoretical ionization potentials of Au$_n$ clusters compare well with experiments. 
\end{abstract}

\pacs{36.40.Cg, 36.40.Mr, 73.22.-f }

\section{Introduction}

Gold clusters are currently attracting interest as the building blocks of novel nanostructured materials and devices. \cite {appl1,appl2,appl3,appl4} During the past two decades, the structures of gold clusters have been studied both experimentally \cite{xrpd,stm,ups,xray,time,optical1} and theoretically. \cite {zhao3,hand,cna,soler,Oliver,gstz,barnett,ecp,model,lee,hann,hen} Experiments suggest that gold nanoclusters with diameters of $1-2$ nm are amorphous. \cite{xrpd,stm} Theoretical results from empirical molecular dynamics simulations or first-principles calculations also support this argument. \cite{gstz,barnett,lee,doye}

In the past decade, the structure and electronic properties of gold clusters have been intensively studied with various theoretical methods. H$\ddot{a}$kkinen {\em et al.} investigated the neutral and anions of Au$_{2-10}$ clusters with local-spin-density approximation \cite{hann}. Gr$\ddot{o}$nbech {\em et al.} compared Au$_2$ to Au$_5$ with spin-polarized Becke-Lee-Yang-Parr (BLYP) functional \cite{hen}. Bravo-P$\acute{e}$rez {\em et al.} investigated small gold clusters up to 6 atoms at {\em ab initio} Hartree-Fock (HF) and post-HF\cite{bravo} level. For the large clusters with up to 147 atoms, H$\ddot{a}$berlen {\em et al.} performed a scalar relativistic all-electron density functional calculations on several magic-sized clusters with icosahedral, octahedral and cuboctahedral structures \cite{Oliver}. By combining pseudopotential DFT calculations with an empirical genetic algorithm, Garz$\acute{o}$n found amorphous structures in Au$_n$ ($n=38,55,75 $) clusters. \cite{gstz} In addition, there are some other works on the global minimal structures of medium-sized gold clusters based on empirical potentials. \cite{lee,emp1,emp2,emp3} 

Despite the achieved progress, there are still many open questions for gold clusters. For example, no direct experimental information is available on the structures of smaller Au$_n$ clusters down to $38$ atoms. Thus, accurate first principles calculation is important for understanting the structural and electronic properties of those clusters. Most previous {\em ab initio} calculations on gold clusters are limited by the presumed symmetric constraints. \cite{xrpd,Oliver} A unconstrained global search on the cluster potential energy surface is needed. \cite{cna} In this work, we generate a number of structural isomers from empirical genetic algorithm simulations. These metastable isomers are further optimized at DFT level to determine the lowest-energy structures. The relative stabilities, electronic density of states, HOMO-LUMO gap and ionization potentials of gold clusters are investigated. 

\section{Computational Method}

In this work, self-consistent field (SCF) electronic structure calculations on gold clusters are performed by using a DFT-based DMol package \cite{dmol}. A relativistic effective core potential (ECP) and a double numerical basis including $d$-polarization function are used in the calculations. The electron density functional is treated by the local density approximation (LDA) with the exchange-correlation potential parameterized by Perdew and Wang \cite{pw}. SCF calculations are carried out with a convergence criterion of 10$^{-6}$ a.u. on the total energy and electron density. Geometry optimizations are performed with the Broyden-Fletcher-Goldfarb-Shanno (BFGS) algorithm. We use a convergence criterion of 10$^{-3}$ a.u on the gradient and displacement, and 10$^{-5}$ a.u. on the total energy in the geometry optimization.

The accuracy of the current computational scheme has been checked by benchmark calculations on the gold atom and the bulk gold solid. The ionization potential and electron affinity obtained for gold atom from our calculation are 9.78 eV and 2.51 eV respectively, which agree well with experimental data: 9.22 eV \cite{kittle} and 2.31 eV \cite{huheey}. For gold solid in fcc structure, we obtained a lattice parameter as 4.07 $\stackrel{\text{o}}{\text{A}}~$ and the cohesive energy as 4.01 eV per atom, while the experimental lattice parameter is 4.08 $\stackrel{\text{o}}{\text{A}}~$ and experimental cohesive energy is 3.81 eV per atom \cite{kittle}. Thus, we believe the effective core potential and numerical basis set used in current DFT package is reasonably good to describe different gold systems from atom to clusters and solid.

Due to the complexity of the electron configuration of gold atoms ($5d^{10}6s^1$), simulated annealing (SA) determination of cluster global minimal structure at DFT level is rather computational expensive. Alternatively, we generate a number of low-energy structural isomers for each cluster size by using a genetic algorithm (GA). \cite{lee,ga1,ga2} In the GA simulation, we adopt molecular dynamics with a properly fitted tight-binding potential to perform local structural relaxtions. \cite{pot} The structures obtained from empirical GA simulations are then fully optimized by LDA calculation to locate the global lowest-energy configuration. The essential idea is to divide the phase space into a number of regions and find a locally stable isomer to represent each region. Our previous works show that the combined scheme of DFT with empirical GA is a reliable and efficient way for modeling the structural properties of atomic clusters up to 25 atoms \cite{zhao2}.

\section{Results and Discussions}

\begin{figure}
\centerline{
\epsfxsize=3.5in \epsfbox{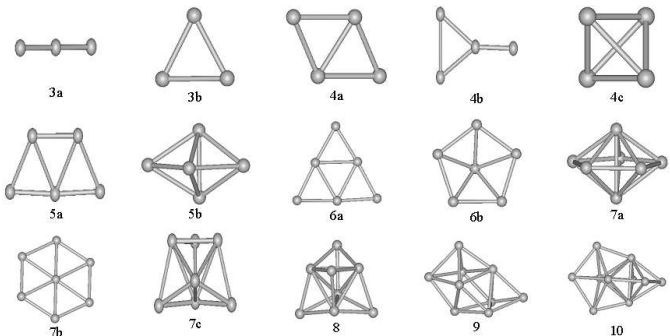}
}
\caption{The lowest-energy and metastable structures for small Au$_n$ clusters: $n=3-10$.}
\end{figure}

The obtained lowest-energy structures and some metastable isomers for Au$_n$ clusters are shown in Fig.1 ($n=3-10$) and Fig.2 ($n=11-20$). For the Au$_2$ dimer, the binding energy, bond length and vibration frequency are obtained as $2.43$ eV, $2.55$ $\stackrel{\text{o}}{\text{A}}$\ and $173$ cm$^{-1}$, respectively. Our current LDA results are in satisfactory agreement with the experimental data ($Eb=2.30$eV, R$_e=2.47\stackrel{\text{o}}{\text{A}}$\ and $\omega =191$cm$^{-1})$. \cite{exp1,exp2} For Au$_3$ trimer, a linear chain (3a) with bond length 2.67 $\stackrel{\text{o}}{\text{A}}$\ is about $0.04$ eV lower in energy than the triangle structures (3b), which is in consistent with CASSCF studies \cite{cass} but on contrary to BLYP results. \cite{hann} Due to the Jahn-Teller instability, both the obtuse and acute triangle of Au$_3$ are more stable than the equilateral triangle, while the energy differences between these triangle isomers are very small, i.e., within $0.01$ eV.

The lowest-energy structures of gold clusters with $4-6$ atoms are found to adopt planar forms. For Au$_4$, a planar rhombus (4a) is about $0.21$ eV in energy lower than the planar ''Y-shaped'' structure (4b) and 1.60 eV lower than the three-dimensional (3D) tetrahedron (4c). For Au$_5$, the trapezoidal (``W-shape'') structure with $C_{2v}$ symmetry (5a) is more stable than the 3D trigonal bipyramid (5b) by 0.79 eV and the square pyramid by 0.94 eV. In the case of Au$_6$, we obtain a planar triangle with $D_{3h}$ symmetry (6a). It can also be understood in terms of ''W-shaped'' Au$_5$ capping by an extra atom. The 3D configurations such as pentagonal pyramid (6b), octahedron and capped trigonal bipyramid are found as local minimum for Au$_6$. The experimental photodetachment spectra of Au$_6^{-}$ cluster also implies a planar hexagonal structure with low electron affinity and large HOMO-LUMO gap. \cite{taylor} The planar equilibrium structures have been obtained for Au$_{4-6}$ from previous {\em ab initio} calculations \cite{hann,hen,bravo}, but can not be obtained from empirical simulations \cite{emp1,emp2}. It is worthy to note that the other small monovalent metal clusters such as Na$_n$\cite{alk}, Ag\cite{Ag} and Cu\cite{Cu} also adopt similar planar configurations.

The pentagonal bipyramid (7a) is the lowest-energy structure found for Au$_7$, which is more stable than distorted capped octahedron by 0.03 eV, planar hexagonal structure (7b) by 0.16 eV, and capped octahedron structure by 0.19 eV. The most stable configurations for Au$_8$ and Au$_9$ are largely distorted bicapped octahedron and bicapped pentagonal bipyramid, respectively. The onset of 3D lowest-energy structures starting from Au$_7$ indicate a 2D$\rightarrow 3$D transition around the size of $7$ atoms. Similar structural changes have been also found in alkali-metal clusters. \cite{alk}

\begin{figure}
\centerline{
\epsfxsize=3.5in \epsfbox{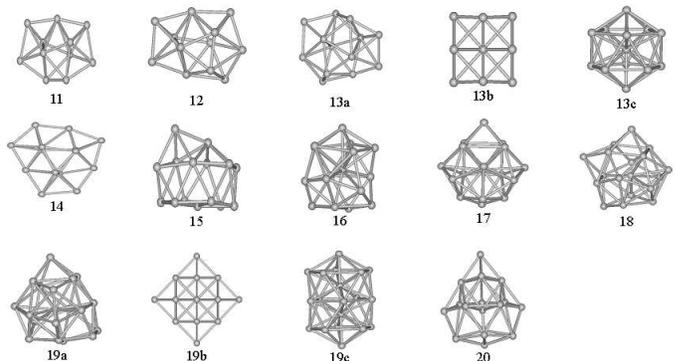}
}

\caption{The lowest-energy and metastable structures for medium-sized Au$_n$ clusters: $n=11-20$.}
\end{figure}

Some structural components related to icosahedron or cuboctahedron, such as pentagonal and hexagonal structures are found in Au$_n$ from $n\geq 10$. For example, both Au$_{10}$ and Au$_{11}$ can be taken as two interpenetrated pentagonal bipyramid, while the ground state structure of Au$_{12}$ is constituted by two hexagons and three pentagons. However, the lowest-energy structure for Au$_{13}$ is neither icosahedron (13c) nor cuboctahedron (13b), but a amorphous configuration (13a in Fig.2). This disordered amorphous configuration is about 1.44 eV in energy lower than cuboctahedron and about 2.71 eV lower than icosahedron. From our calculations, the cuboctahedron is more stable than icosahedron, which has been predicted by H$\ddot{a}$berlen. \cite{Oliver} We have also examined the bond lengths distributions of different isomers for Au$_{13}$. The disordered isomer shows broader bond length distributions than those for the high symmetric structures. Similar results were found in Pt$_{13}$ cluster from previous LDA calculation. \cite{Pt} 

Tabular cage structures are found for the Au$_n$ clusters with $n=10-14$. Especially, Au$_{14}$ can be taken as three interpenetrated pentagonal bipyramid. This behavior is quite different from those in the silver and copper clusters, whose ground state configurations are usually near-spherical icosahedron-like structures \cite{Ag,Cu,zhao1}. In the case of Au$_{15}$, we find the characteristics belong to both tabular cage and compact structure, implying a transition from tabular geometry to compact one. As $n\geq 16$, the compact near-spherical structures become dominant and the structures can be obtained by capping on distorted icosahedron. However, despite there are some icosahedral-like features in Au$_{19}$, the double icosahedron is not found as lowest-energy structure. Similar to Au$_{13}$, the amorphous structure (19a) is the most stable configuration for Au$_{19}$. The energy difference between amorphous and fcc structure (19b) is $1.95$ eV and the difference between the amorphous and the double icosahedron (19c) is 2.83 eV. Since the amorphous structures are also found for larger Au$_{n}$ ($n=38, 55, 75$) \cite{gstz}, we suggest that the amorphous packing is the common structural feature of Au clusters. By using the same computational scheme, we have studied the structures properties of larger gold cluster within the size range $n=20-38$. Amorphous structures are also found and the detailed results will be presented in a forthcoming paper.

\begin{figure}
\vspace{0.65in}
\centerline{
\epsfxsize=3.0in \epsfbox{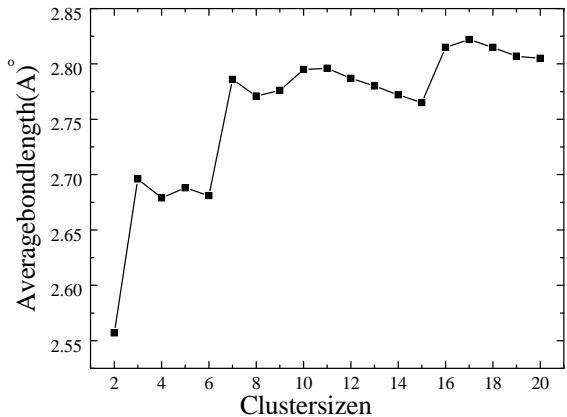}
}
\vspace{-0.75in}
\caption{The average bond length of Au$_n$ versus cluster size.}
\end{figure}

To further elucidate the growth trend of Au clusters and the size evolution of Au-Au interactions, we describe the average bond length as a function of cluster size in Fig.3. Three distinct steps corresponding to planar structures, tabular structures and compact structures are found. The twice increases of the average bond length in Fig.3 reflect the structural transitions at $n=7$ and $n=15$. The planar structure can be understood by the directionality of Au-Au bonds, implying some covalent feature in small Au clusters. As the cluster size increases, metallic bonding characters become important, which leads to the coexistence of planar structural and compact structural characters. Thus, the tabular structures are preferred in the Au$_{10-14}$ clusters. As the cluster size further increases, metallic bonding eventually prevail. Therefore, compact structure appears as the dominating growth pattern of the medium-size gold clusters.

The appearance of different structural characters in gold clusters can be understood by the interplay between $5d$ and $6s$ electrons. In the small gold clusters, the 6$s$ valence electrons are predominant, while the $5d$ states are low-lying and contribute less to the electronic behavior. Thus, the small gold clusters should exhibit certain alkali-metal-like behaviors, e.g., planar ground state configuration, which can be described by $s$-orbital modified H$\ddot{u}$ckel model. \cite{zhao3} As the cluster size increases, the $sd$ hybridization increase and the contributions from $d$ electrons become more important. Thus the clusters tend to adopt the more compact structures. However, as the short-range interaction between gold atoms is extraordinary strong, it favors a tabular structure in medium-sized range. The discussions below show that the $6s$ electrons still plays significant role in determining the electronic properties such as IPs, HOMO-LUMO gaps.

\begin{table}[tbp]
Table I. Lowest-energy structures and electronic properties of Au$_n$ clusters. $E_b$ (eV): bind energy per atom; $\Delta$ (eV): HOMO-LUMO gap; IP$^a$ (eV): theoretical adiabatic ionization potentials; IP$^b$ (eV): experimental ionization potentials \cite{5}.
\par
\begin{center}
\begin{tabular}{cccccc}
$n$ & Geometry                     & $E_b$ & $\Delta$ & IP$^a$ & IP$^b$ \\ \hline
  2 & dimer                        & 1.22 & 1.94 & 9.81 & 9.50 \\ 
  3 & linear chain & 1.28 & 2.70 & 7.30 & 7.50 \\ 
  4 & rhombus & 1.74 & 1.02 & 8.34 & 8.60 \\ 
  5 & planar trapezoid (W-form) & 1.90 & 1.51 & 7.78 & 8.0 \\ 
  6 & planar triangle & 2.18 & 2.06 & 8.55 & 8.80 \\ 
  7 & pentagonal bipyramid & 2.13 & 1.00 & 7.20 & 7.8 \\ 
  8 & distorted bicapped octahedron & 2.30 & 2.09 & 8.19 & 8.65 \\ 
  9 & bicapped pentagonal bipyramid & 2.30 & 0.97 & 7.22 & 7.15 \\ 
 10 & tabular structure & 2.39 & 1.03 & 7.35 & 8.2 \\ 
 11 & tabular structure & 2.42 & 0.86 & 7.20 & 7.28 \\ 
 12 & tabular structure & 2.50 & 0.82 & 7.55 & 8.15 \\ 
 13 & tabular structure & 2.53 & 0.63 & 6.84 & 7.70 \\ 
 14 & tabular structure & 2.62 & 1.58 & 7.65 & 8.00 \\ 
 15 & tabular and compact structure & 2.60 & 0.22 & 7.04 & 7.65 \\ 
 16 & compact structure & 2.63 & 0.44 & 7.38 & 7.80 \\ 
 17 & compact structure & 2.69 & 0.81 & 7.33 & 7.60 \\ 
 18 & compact structure & 2.75 & 0.92 & 7.44 & 7.85 \\ 
 19 & compact structure & 2.77 & 0.70 & 6.82 & 7.70 \\ 
 20 & compact structure & 2.79 & 0.77 & 7.05 & 7.82
\end{tabular}
\end{center}
\end{table}

Based on the lowest-energy structures, we now discuss the electronic properties of gold clusters. In this work, we have also calculated the adiabatic ionization potentials (IPs) from the total energy difference between the ground state neutral Au$_n$ and the fully relaxed cationic Au$_n^{+}$ clusters. The binding energy, the gap between highest occupied molecular orbital (HOMO) and lowest occupied molecular orbital (LUMO), and ionization potentials for Au$_n$ clusters are presented in Table I along with experimental IPs \cite{5}. As shown in Table I, for small clusters, the binding energy increase rapidly with cluster size. As $n\geq 8$, the size dependent increase of binding energy become slower, corresponding to the metallic cohesion in the Au$_n$ clusters. It is worthy noted that the binding energy up to Au$_{20}$ is only 2.79 eV, which is about 70$\%$ of the bulk cohesive energy. 

\begin{figure}
\vspace{-0.15in}
\centerline{
\epsfxsize=3.0in \epsfbox{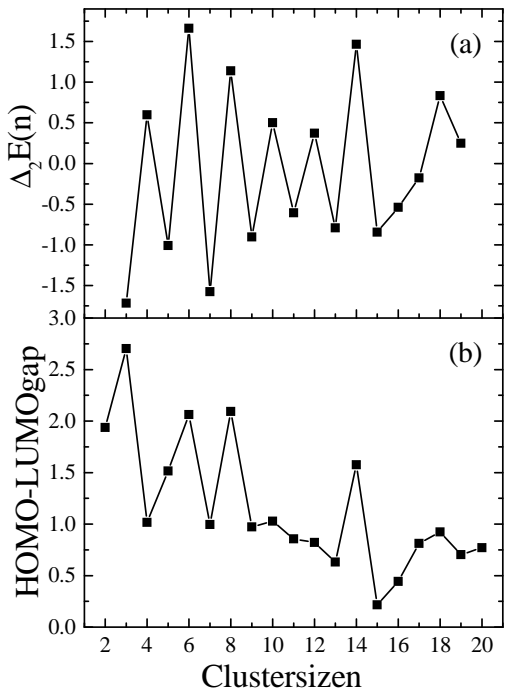}
}
\vspace{-0.25in}
\caption{Odd-even oscillations of cluster properties with cluster size: (a) Second differences of cluster energies $\Delta _2E(n)=E(n-1)+E(n+1)-2E(n)$ (eV); (b) HOMO-LUMO gaps (eV).}
\end{figure}

In Fig.4a and 4b, we plot the second differences of cluster total energies $\Delta _2E(n)=E(n+1)+E(n-1)-2E(n)$ and HOMO-LUMO gaps as a function of cluster size. Both the $\Delta _2E(n)$ and HOMO-LUMO gap exhibit dramatic odd-even oscillations. The even-numbered Au$_n$ clusters are relatively more stable than the neighboring odd-sized ones and have larger HOMO-LUMO gap. The odd-even oscillation behaviors were observed experimentally \cite{ups,5} and can be understood by electron pairing effect. Odd(even)-sized clusters have an odd(even) total number of $s$ valence electrons and the HOMO is singly(doubly) occupied. The electron in a doubly occupied HOMO feels a stronger effective core potential since the electron screening is weaker for the electrons in the same orbital than for inner shell electrons. Therefore, the binding energy of a valence electron in a cluster of even size cluster is larger than that of odd one. In Fig.4b, the HOMO-LUMO gap for Au$_2,$ Au$_6$ and Au$_8$ is particularly large (1.94 eV, 2.06 eV, 2.09 eV), which compare well with previous calculations by H$\ddot{a}$kkinen (1.96 eV, 2.05 eV and 2.04 eV). \cite{hann} For $n=10-20$, substantial HOMO-LUMO gap is found in the Au$_{10}$, Au$_{14}$ and Au$_{18}$ clusters (1.03 eV, 1.58 eV and 0.92 eV). The high stability of Au$_8$ and Au$_{18}$ can be understood by the effect of $s$-electron shell, which is also found in silver clusters \cite{zhao1}. The extraordinary tabular configurations for the Au$_n$ clusters with $n=10-14$ might be attributed to the interplay between electronic and geometric effect.

To explore the geometric effect on the electronic structure, we compare the electronic density of states (DOS) of Au$_{13}$ for the three isomers: amorphous; cuboctahedron; icosahedron in Fig.5. The cluster electronic DOS shows remarkable structural sensitivity. The DOS for amorphous structures demonstrate more uniform distributions in comparison with the other two cases, because of the lower symmetry. Such structural dependence of electronic state can be used to identify the cluster geometries with the aid of experimental spectroscopy measurements. 

\begin{figure}
\vspace{-0.25in}
\centerline{
\epsfxsize=3.0in \epsfbox{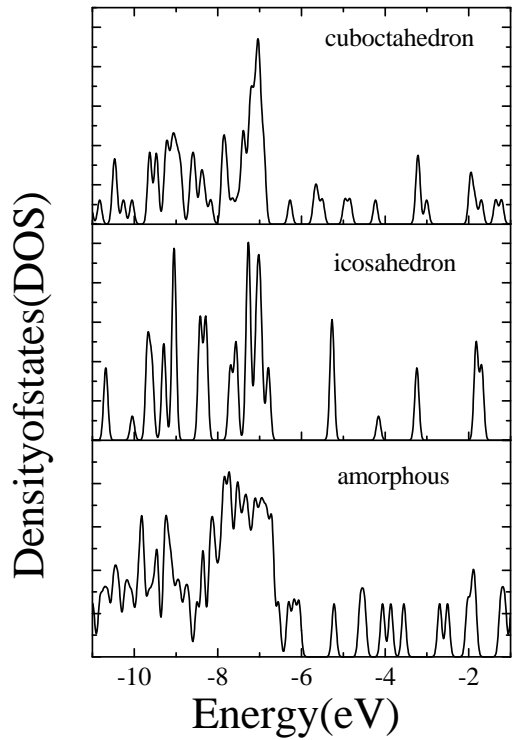}
}
\caption{Density of states (DOS) of the Au$_{13}$ clusters with different geometries: cuboctahedron, icosahedron, amorphous. Gaussian broadening of 0.05eV is used.}
\end{figure}

We further investigate the size evolution of electronic properties of gold clusters by examining the electronic density of states (DOS) of several representative clusters: Au$_2$, Au$_8$, Au$_{18}$. As shown in Fig.6, in smallest clusters Au$_2$, the energy levels are discrete and $d$ and $sp$ peaks are clearly separated. The DOS of Au$_8$ is still sparse and discrete although the $d$ and $sp$ energy levels are gradually broadened. As the cluster size further increases, the $d$ and $sp$ levels broaden, shift and overlap with each other. Thus, continuous electronic bands are found in Au$_{18}$. In a previous experimental study of Au$_n$ clusters up to 223 atoms \cite{ups}, the ultraviolet photoelectron spectra of smallest Au$_n$ ($n\leq 10\sim 13$) depends sensitively on the cluster size. The size evolution of photoelectron spectra for Au$_n$ with $n>12$ becomes more gradual and the spectra of Au$_{80}$ is already quite similar to that of solid silver. 

\begin{figure}
\vspace{-0.25in}
\centerline{
\epsfxsize=3.0in \epsfbox{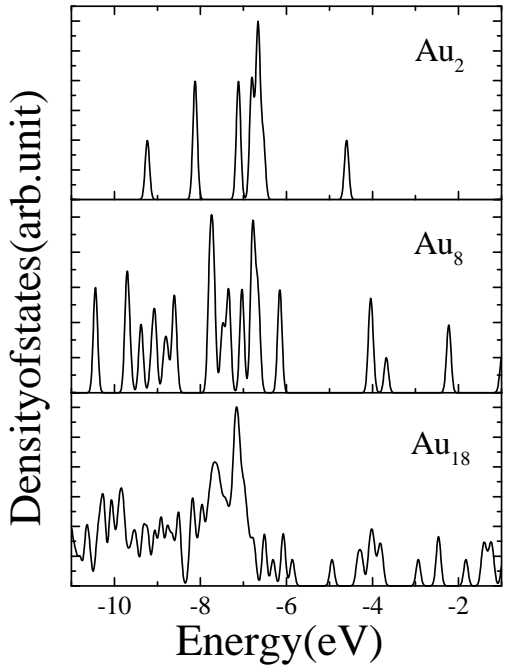}
}
\caption{Density of states (DOS) of the Au$_2$, Au$_8$, and Au$_{18}$ clusters. Gaussian broadening of 0.05eV is used.}
\end{figure}

In cluster physics, the ionization potential is an important property that reflects the size-dependent evolution of electronic structure. For alkali clusters such as Na$_n$, K$_n$, the IPs converge to its bulk limit (work function of solid) linearly with $n^{-1/3}$ (or $1/R$, $R$ is the cluster radius) \cite{2}. This behavior can be understood by a conducting spherical droplet (CSD) model \cite{19,20}. In Fig.7, the theoretical and experimental IPs of Au$_n$ are plotted as a function of $n^{-1/3}$ and compared with the prediction from classical conducting sphere droplet (CSD) model \cite{20}. Our present results agree well with the experiments, while the CSD model can only qualitatively describe the size dependence of IPs. Similar to those in Fig.4, dramatic even-odd alternative behavior is found in Fig.7, where clusters with even number of $s$ valence electrons have higher IPs than their immediate neighbors. In addition, particular higher IP values at the magic-sized clusters such as Au$_2$, Au$_4$, Au$_{6\text{,}}$ Au$_8$, Au$_{14}$ and Au$_{18}$ is obtained. Some of the magic size (Au$_2$, Au$_8$, Au$_{14},$ Au$_{18}$) can be associated to the occupation of electronic shell \cite{2}. The IP behavior found for Au$_n$ clusters is very similar to those in Ag$_n$ clusters \cite{zhao2}. The significant deviation from the CSD model demonstrate that the Au$_n$ clusters up to $n=20$ is still far from a piece of bulk metal.

\begin{figure}
\vspace{0.75in}
\centerline{
\epsfxsize=3.0in \epsfbox{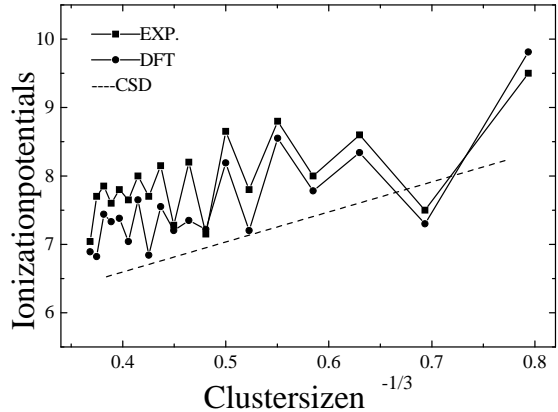}
}
\vspace{-0.5in}
\caption{Ionization potentials (IPs) of Au$_n$. Squares: experimental data \cite{5}; circles: our DFT calculation; dashed line: CSD model \cite{20}.}
\end{figure}

\section{Conclusion}

In summary, the lowest-energy geometries, binding energies, density of states, HOMO-LUMO gap and ionization potentials of Au$_n$($n=2-20$) clusters have been obtained by LDA calculations combined with empirical genetic algorithm simulations. The main conclusions can be made as follows. (1) The structures of smallest gold clusters are planar and dominated by $s$ electrons, similar to those of alkali-metal clusters and other coinage-metal clusters. The contributions of $d$ electrons become more important and the structural transition from 2-D to 3-D takes place at the size of 7 atoms. (2) The electronic effect and geometrical effect simultaneously influence the ground state configurations of medium-sized clusters, which leads to the tabular cage configurations with $n=10-14$ and more compact structures with $n\geq 15$. (3) The most stable configurations of Au$_{13}$ and Au$_{19}$ are neither icosahedron nor cuboctahedron, but amorphous structures. Remarkable difference in electronic states are found between the structural isomers. (4) The odd-even alternation behaviors are found in the relative stabilities, density of states, HOMO-LUMO gap and ionization potentials of gold clusters. The even-numbered Au$_n$ clusters show relatively higher stability. Bulk-like continuous electronic band is found in Au$_{18}$ while the electronic behavior of such clusters is still far from bulk metal.

\begin{acknowledgements}

The authors would like to thank financial support from National Nature Science Foundation of China (No.29890210), the U.S. ARO (No.DAAG55-98-1-0298), and NASA Ames Research Center. We acknowledge the computational support from the North Carolina Supercomputer Center.
\end{acknowledgements}

\ \newline
$^{*}$ Email: wangqun@nju.edu.cn. \newline
$^{\dagger}$ Email: zhaoj@physics.unc.edu.


\begin{references}

\bibitem{appl1}  R.L.Whetten, J.T.Khoury, M.M.Alvarez, S. Murthy, I. Vez-mar,Z. L. Wang, P. W. Stephens, C. L. Cleveland, W. D. Luedtke, and U. Landman, Adv. Mater. {\bf 5}, 8(1996).

\bibitem{appl2}  R.P.Andres, T. Bein, M. Dorogi, S. Feng, J.I.Henderson, C.P.Kubiak, W. Mahoney, R.G.Osifchin and R. Reifenberger, Science {\bf 272}, 1323(1996).

\bibitem{appl3}  C.A.Mirkin, R.L.Letsinger, R.C.Mucic, J. Storhoff, Nature {\bf 382}, 607(1996).

\bibitem{appl4}  A.P.Alivisatos, K.P.Johnsson, X.Peng, T.E. Wilson, C.J.Loweth, M.P. Bruchez, P.G. Schultz , Nature {\bf 382}, 609(1996).

\bibitem{xrpd}  C.L.Cleveland, U.Landman, T.G.Schaaff, M.N. Shafigullin, P.W. Stephens, and R.L. Whetten, Phys. Rev. Lett.{\bf 79}, 1873(1997).

\bibitem{stm}  T.G.Schaaff, W.G.Cullen, P.N.First, I.Vezmar, R. L. Whetten, W. G. Cullen, P. N. First, C. Guti$\acute{e}$rrez-Wing, J. Ascensio, M. J. Jose-Yacam$\acute{a}$n, J. Phys. Chem. {\bf 101} 7885(1997).

\bibitem{ups}  K.J. Taylor, C.L.Pettiette-Hall, O.Cheshnovsky, R. E. Smalley, J. Chem. Phys.{\bf 96}, 3319(1992).

\bibitem{xray}  K.Koga, H. Takeo, T. Ikeda, K.I.Ohshima, Phys.Rev. B{\bf 57}, 4053(1998).

\bibitem{time}  V.A.Spasov, Y.Shi, K.M.Ervin, Chem.Phys.{\bf 262} ,75(2000).

\bibitem{optical1} B.Palpant, B.Prevel, J.Lerme, E.Cottancin, M.Pellarin, M.Treilleux, A.Perez, J.L.Vialle,M.Broyer, Phys.Rev.B{\bf 57}, 1963(1998).

\bibitem{zhao3}  J.J.Zhao, X.S.Chen, G.H.Wang, Phys.Lett.A{\bf 189}, 223(1994).

\bibitem{hand}  H.Handschuh, G.Gantef$\ddot{o}$r, P.S.Bechthold, W.Eberhardt, J. Chem.Phys.{\bf 100}, 7093(1994).

\bibitem{cna}  I.L.Garzon, A.Posada-Amarillas, Phys. Rev.B{\bf 54}, 11796(1996).

\bibitem{soler}  J.M.Soler, M.R.Beltran, K.Michaelian, I.L.Garzon, P.Ordejon, D.Sanchez-Portal, E. Artacho, Phys.Rev.B {\bf 61}, 5771(2000).

\bibitem{Oliver}  O.D. H$\ddot{a}$berlen, S.C.Chung, M.Stener, N. R$\ddot{o}$sch, J.Chem.Phys.{\bf 106}, 5189(1997).

\bibitem{gstz}  I.L.Garz$\acute{o}$n, K.Michaelian, M.R.Beltr$\acute{a}$n, A. Posada-Amarillas, P. Ordej$\acute{o}$n, E. Artacho, D.S$\acute{a}$nchez-Portal, and J. M. Soler, Phys.Rev.Lett. {\bf 81}, 1600(1998).

\bibitem{barnett}  R.N.Barnett, C.L. Cleveland, H.H$\ddot{a}$kkinen, W.D.Luedtke, C. Yannouleas, and U. Landman, Eur.Phys.J.D {\bf 9}, 95(1999).

\bibitem{ecp}  J.L.BelBruno, Heteroatom Chem {\bf 9}, 651(1998).

\bibitem{model}  N.T. Wilson, R.L. Jphnston, Eur.Phys. J.D {\bf 12}, 161(2000).

\bibitem{lee}  T.Li, S.Yin, Y.Ji, G.Wang, J.Zhao, Phys.Lett.A{\bf 267}, 403(2000).

\bibitem{hann}  H.H$\ddot{a}$kkinen and U. Landman, Phys.Rev.B{\bf 62}, 2287(2000).

\bibitem{hen}  H. Gr$\ddot{o}$nbech, W. Andreoni, Chem.Phys.{\bf 262}, 1(2000).

\bibitem{doye}  J.P.K. Doye, D.J.Wales, New J.Chem.{\bf 22} 733(1998).

\bibitem{bravo}  G.Bravo-P$\acute{e}$rez, I.L.Garz$\acute{o}$n and O.Novaro, J.Mol. Stru.(Theochem) {\bf 493}, 225(1999).

\bibitem{emp1}  K.Michaelian, N.Rendon, and I.L.Garzon, Phys.Rev.B{\bf 60}, 2000(1999).

\bibitem{emp2}  N.T.Wilson and R.L.Johnston, Eur.Phys.J.D{\bf 12}, 161(2000).

\bibitem{emp3}  J.M.Soler, I.L.Garz$\acute{o}$n, J.D.Joannopoulos, Solid State Commun.{\bf 117}, 621(2001).

\bibitem{dmol} DMOL is a density functional theory (DFT) based package with atomic basis distributed by Accelrys (B.Delley, J.Chem.Phys.{\bf 92}, 508(1990). 

\bibitem{pw}  J.P.Perdew and Y.Wang, Phys.Rev.B{\bf 45}, 13244(1992).

\bibitem{kittle}  C.Kittle, {\em Introduction to Solid State Physics}, 7th edition, (John Wiley $\&amp;$ Sons, New York, 1996).

\bibitem{huheey}  J.E.Huheey, E.A.Keiter, and R.L.Keiter, {\em Inorganic Chemistry: Principles of Structure and Reactivity}, 4th edition, (HarperCollins, New York, 1993).

\bibitem{ga1}  D.M.Deaven, K.M.Ho, Phys.Rev.Lett.{\bf 75}, 288(1995). 

\bibitem{ga2}  B.Wang, S.Yin, G.Wang, A.Buldum, and J. Zhao, Phys.Rev.Lett.{\bf 86}, 2046(2001); Y.H.Luo, J.J.Zhao, S.T.Qiu, G.H.Wang, Phys.Rev.B{\bf 59},  14903(1999).

\bibitem{pot}  F.Cleri, V.Rosato, Phys.Rev.B {\bf 48}, 22(1993).

\bibitem{zhao2} J.L.Wang, G.H.Wang, J.J.Zhao, Phys.Rev.B{\bf 64}, 205411(2001); J.Phys.Conden.Matter{\bf 13}, L753(2001); J.J.Zhao, Phys.Rev.A{\bf 64}, 043204(2001).

\bibitem{exp1}  {\em CRC Handbook of Chemistry and Physics}, 55th ed., edited by R.C. Weast (CRC Press, Cleveland, Ohio, 1974).

\bibitem{exp2}  {\it American Institute of Physics Handbook}, (McGraw-Hill, New York, 1972).

\bibitem{cass}  K. GBalasubramanian, M.Z Liao, J.Chem.Phys. {\bf 86}, 5587(1987).

\bibitem{taylor}  K.J.Taylor, C.Jin, J.Conceicao, O. Cheshnovsky, B. R.Johnson, P.J. Nordlander, and R. E. Smalley, J.Chem.Phys.{\bf 93} 7515(1991).

\bibitem{alk}  U.R$\ddot{o}$thlisberger and W. Andreoni, J.Chem.Phys.{\bf 94}, 8129(1991).

\bibitem{Ag}  V. Bonaci-Kouteck, L. Cespiva, P. Fantucci, and J. Koutecky, J.Chem.Phys.{\bf 98} 7981(1993), {\bf 100} 940(1994).

\bibitem{Cu}  C.Massobrio, A. Pasquarello, A. Dal Corso, J.Chem. Phys.{\bf 109}, 6626(1998). 

\bibitem{Pt}  S.H.Yang, D.A. Drabold, J.A. Adams, P. Ordejon, K. Glassford, J.Phys.Cond. Matter {\bf 9}, L39 (1997).

\bibitem{zhao1}  J.J.Zhao, Y.H.Luo, G.H.Wang, Euro.J.Phys.D{\bf 14}, 309(2001).

\bibitem{5}  C.Jackschath, I.Rabin, W.Schulze, Ber. Bunsenges,Phys.Chem.{\bf 86}, 1200(1992).

\bibitem{2}  W.A.de Heer, Rev.Mod.Phys.{\bf 65}, 611(1993).

\bibitem{19}  D.M.Wood, Phys.Rev.Lett.{\bf 46}, 749(1981).

\bibitem{20}  J.P.Perdew, Phys.Rev.B{\bf 37}, 6175(1988).
\end{references}
\end{document}